\documentclass[a4paper,11pt]{article}
\pdfoutput=1 % if your are submitting a pdflatex (i.e. if you have
\usepackage{jinstpub} % for details on the use of the package, please
\usepackage{graphicx}
\usepackage{subfigure}

\title{\boldmath Laser Calibration System in JUNO}

\author[a]{Yuanyuan Zhang,}
\author[a,1]{Jianglai Liu, \note{Corresponding author.}}
\author[b,c,2]{Mengjiao Xiao, \note{Corresponding author.}}
\author[a]{Feiyang Zhang,}
\author[a]{Tao Zhang}

\affiliation[a]{INPAC and Department of Physics and Astronomy, Shanghai Jiao Tong University, \\Shanghai Laboratory for Particle Physics and Cosmology, Shanghai 200240, China}
\affiliation[b]{Department of Physics, University of Maryland, \\College Park, Maryland 20742, USA}
\affiliation[c]{Center of High Energy Physics, Peking University, \\Beijing 100871, China}

\emailAdd{jianglai.liu@sjtu.edu.cn}
\emailAdd{mxiao@umd.edu}

\abstract{
  A UV laser calibration system is developed for the Jiangmen Underground Neutrino Observatory (JUNO), a 20-kton liquid scintillator-based, long baseline reactor anti-neutrino experiment. This system is capable of delivering fast laser pulses ($<$1~ns) into the detector with a photon intensity ranging from hundreds of keV to a few TeV equivalent energy. An independent intensity monitoring system is developed for this laser, which has achieved a $<$ 0.3\% relative uncertainty. Such a system will be used to calibrate the response of the JUNO photosensors and electronics {\it in situ} to a sub-percent precision. }


\begin{document}
\maketitle
\flushbottom

\section{Introduction}
The Jiangmen Underground Neutrino Observatory (JUNO)~\cite{ref:JUNO_CDR} is a 20-kton multi-purpose liquid scintillator (LS) detector, located 53-km~\cite{ref:baseline} from the Yangjiang and Taishan nuclear power plants in South China. Its primary physics goal is to determine the neutrino mass hierarchy (MH) via the measurement of neutrino energy spectrum with an unprecedented precision. The LS is confined by an acrylic sphere with an inner diameter of 35.7~m, outside of which there is an ultrapure water shielding. About 17000 20-inch photomultiplier tubes (PMTs) and 25000 3-inch PMTs, with maximum quantum efficiency at 420 nm that matches the fluorescence spectrum of the LS, are located in the water viewing the LS. To ensure the MH sensitivity, a sub-percent determination of the energy scale of the LS detector between 1 to 8 MeV is required~\cite{ref:JUNO}, which poses a challenge to the calibration.

The neutrino energy is measured by the prompt positron signal through the so-called inverse-$\beta$ decay, which combines the kinetic energy of the positron and the annihilation gamma rays. The detector energy response can be separated into a physical response which converts the energy deposition to the scintillation and the Cherenkov photons, and an instrumental response which converts optical photons to the PMT signals~\cite{ref:light_propagation}. Radioactive sources calibrate the combined effects, but it is desired to calibrate the instrumental response directly. A laser injection system is an ideal tool for this, which combines a large dynamic range, point-like photon emitting vertex, and fast timing. In this paper, we discuss the development and performance of such a system for JUNO. The rest of this paper is organized as follows. The hardware requirements, design, and experimental setup are discussed in Sec.~\ref{sec:hard}. The performance achieved in the laboratory test will be presented in Sec.~\ref{sec:meas}, followed by the conclusion in Sec.~\ref{sec:con}.

\section{The laser system}
\label{sec:hard}
\subsection{Requirements}
A laser system is standard to calibrate timing and energy response for particle detectors (for previous applications, see refs.~\cite{ref:system_chosen, ref:online_calibration, ref:g2_calibration, ref:g2_light, ref:Super-K, ref:TCal}). In JUNO, the laser system has to satisfy the following requirements:
\begin{enumerate}
\item pulsed beam with $<$ 1~ns timing;
\item the dynamic range of intensity should be equivalent to deposited energies between hundreds of keV (radioactivity events) to a few TeV (showering muon events);
\item isotropic in all directions;
\item producing photons with a similar wavelength and time structure as real physical events;
\item in the anti-neutrino energy range (1 to 8~MeV), the relative laser intensity has to be known to a sub-percent precision;
\item deployable along the central axis of the JUNO detector.
\end{enumerate}

The first and second requirements can be satisfied by a pulsed laser coupled to an optical fiber with the open end deployed into the detector. The third requirement can be fulfilled by adding a PTFE diffuser ball to the open end of the optical fiber. The fourth requirement can be achieved by selecting a laser with ultraviolet (UV) wavelength, whose photons can be absorbed quickly in the LS and re-emitted at a blue wavelength isotropically. The fifth requirement is the major driver of the design, since the laser internal power meter does not have the desired accuracy. Therefore, the laser has to be split, with one beam injected into the detector and the other illuminating a dedicated monitor device with excellent linearity. To satisfy the last requirement, the photons will be transmitted through a shielded optical fiber, deployable by a precisely controlled spool system.

\subsection{Conceptual design}
\begin{figure*}[ht]
\subfigure[Schematics of the laser calibration system]{
  \includegraphics[width=2.7in]{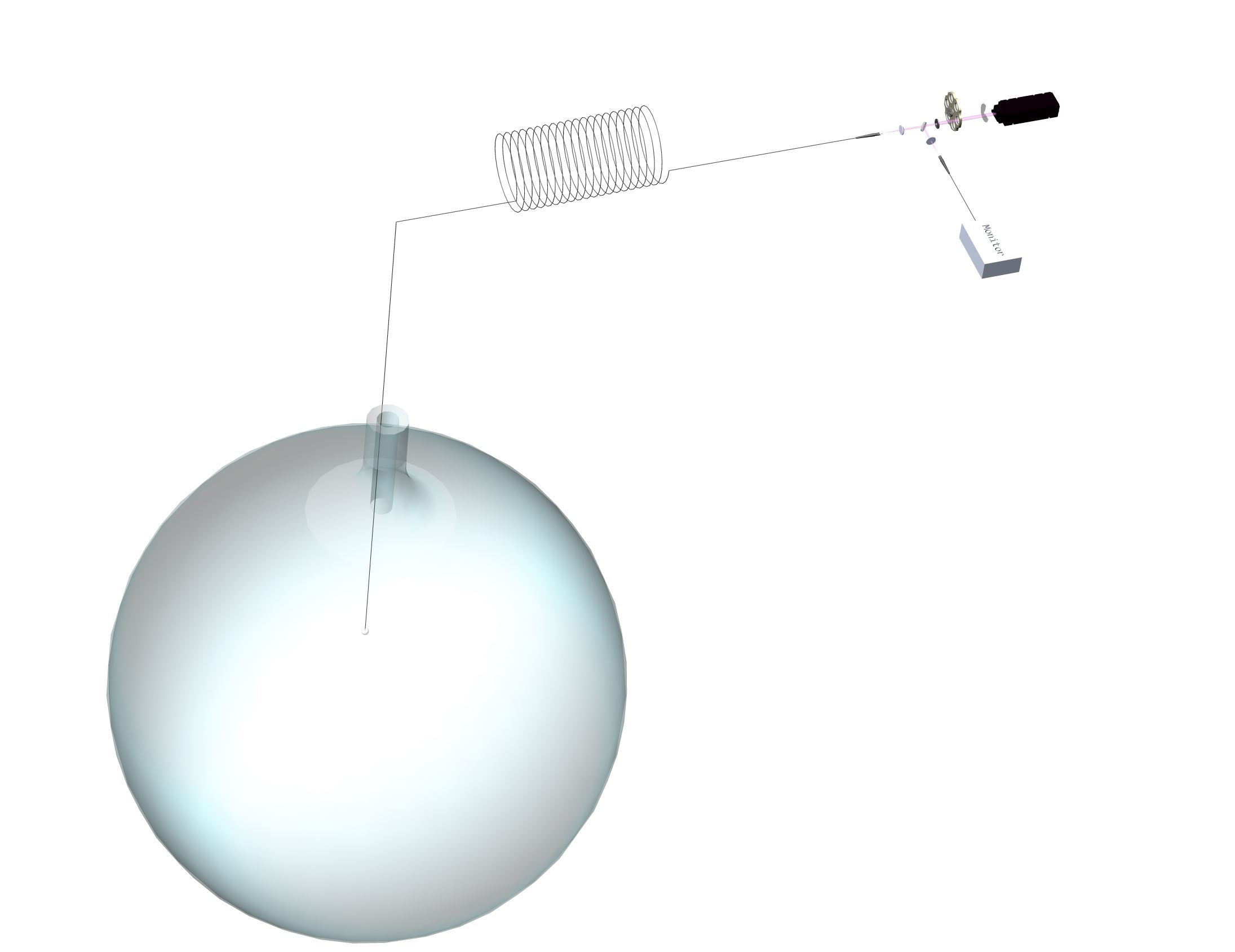}}
  \hspace{0.5in}
\subfigure[a zoom-in of the optical path]{
  \includegraphics[width=2.5in]{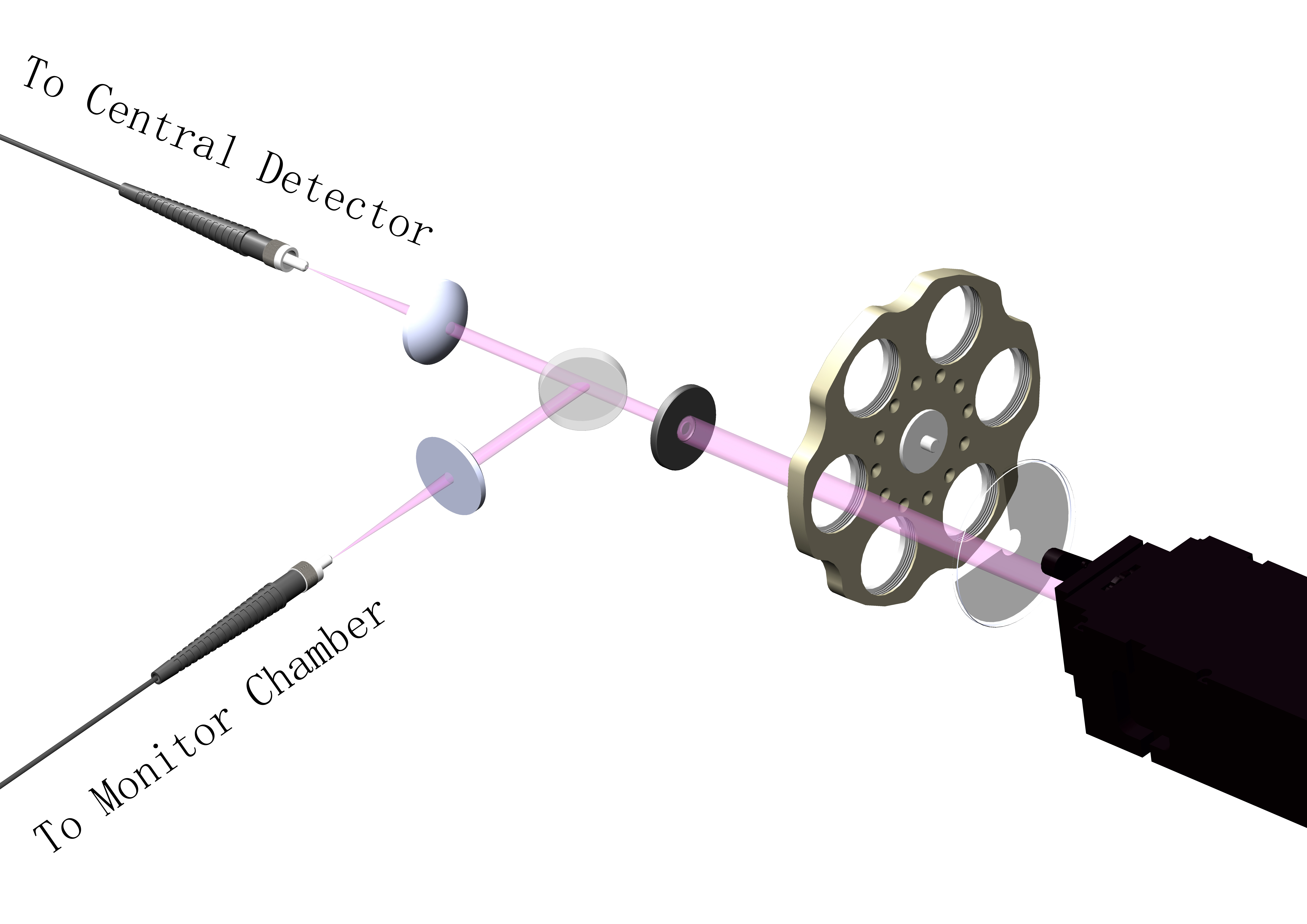}}
\caption{Conceptual design of the laser calibration system.}
\label{concept}
\end{figure*}

The conceptual design of the laser system is shown in Fig.~\ref{concept}. The laser beam goes through a pin hole, then is split into two beams and focused into two fibers. The shorter fiber with 5~m length will bring the photons into a dark box with photomultipliers, which serves as the monitor. The longer fiber with 55~m length will be deployed into the central detector. Assuming that the splitting is stable, and a monitor with perfectly linear response, the ratio between the signals detected by the JUNO detector and the monitor will calibrate the instrumental linearity of the JUNO detector.

\subsection{Laboratory setup}
\begin{figure*}[!htbp]
\centering
\includegraphics[width=1.0\textwidth]{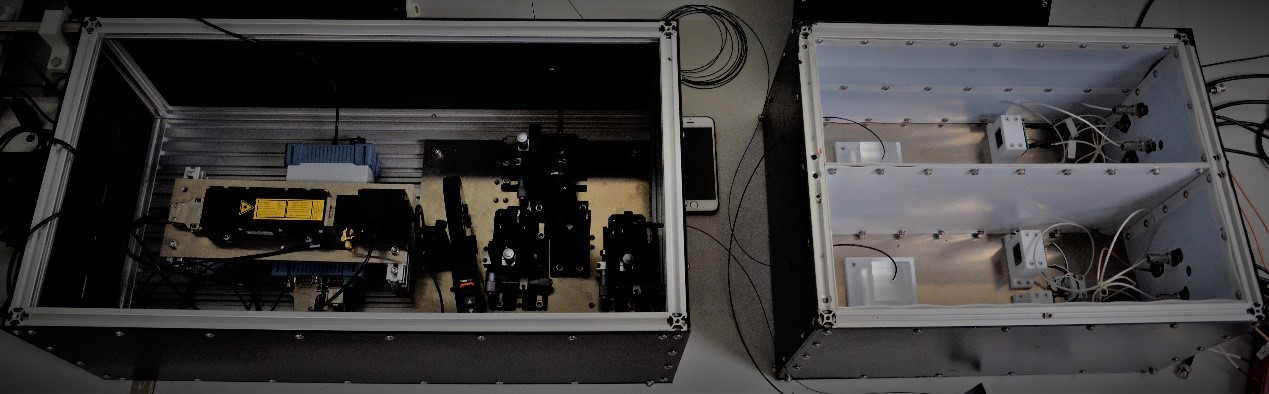}
\caption{Picture of the prototype system, left: laser box, right: dark
  box with monitor and detector chambers.}
\label{system}
\end{figure*}

To demonstrate the feasibility of the system, we have developed a prototype in the laboratory, which contains a laser box to produce the laser beam and another dark box to detect/monitor laser photons.  The laser box contains the UV laser and relevant optical devices. The dark box contains two identical and optically separated chambers, the ``monitor chamber'' and the ``detector chamber'', the former being the laser intensity monitor and the latter simulating the JUNO central detector. The laser is split into two beams and then guided by the optical fibers, one 5-m long into the monitor chamber and the other 55-m long (5-m + optical fiber rotary joint + 50-m, see later) into the detector chamber. The open end of each fiber is inserted into a liquid scintillator container to produce isotropic blue photons, which are viewed by PMTs. Fig.~\ref{system} shows the overall setup of the prototype system. We shall discuss each key component in the following section.

\subsubsection{Fast UV laser}
The JUNO LS uses the same recipe as that for Daya Bay except gadolinium loading~\cite{ref:DYB_LS_Orig, ref:DoubleChooz_FDR}. As discussed in ref.~\cite{ref:LS_spectrum}, a laser with wavelength less than 310~nm is sufficient to produce the same blue emission spectrum centered at 420~nm, similar to that from a real physical event.

\begin{figure*}[!htbp]
\centering
\includegraphics[width=1.0\textwidth]{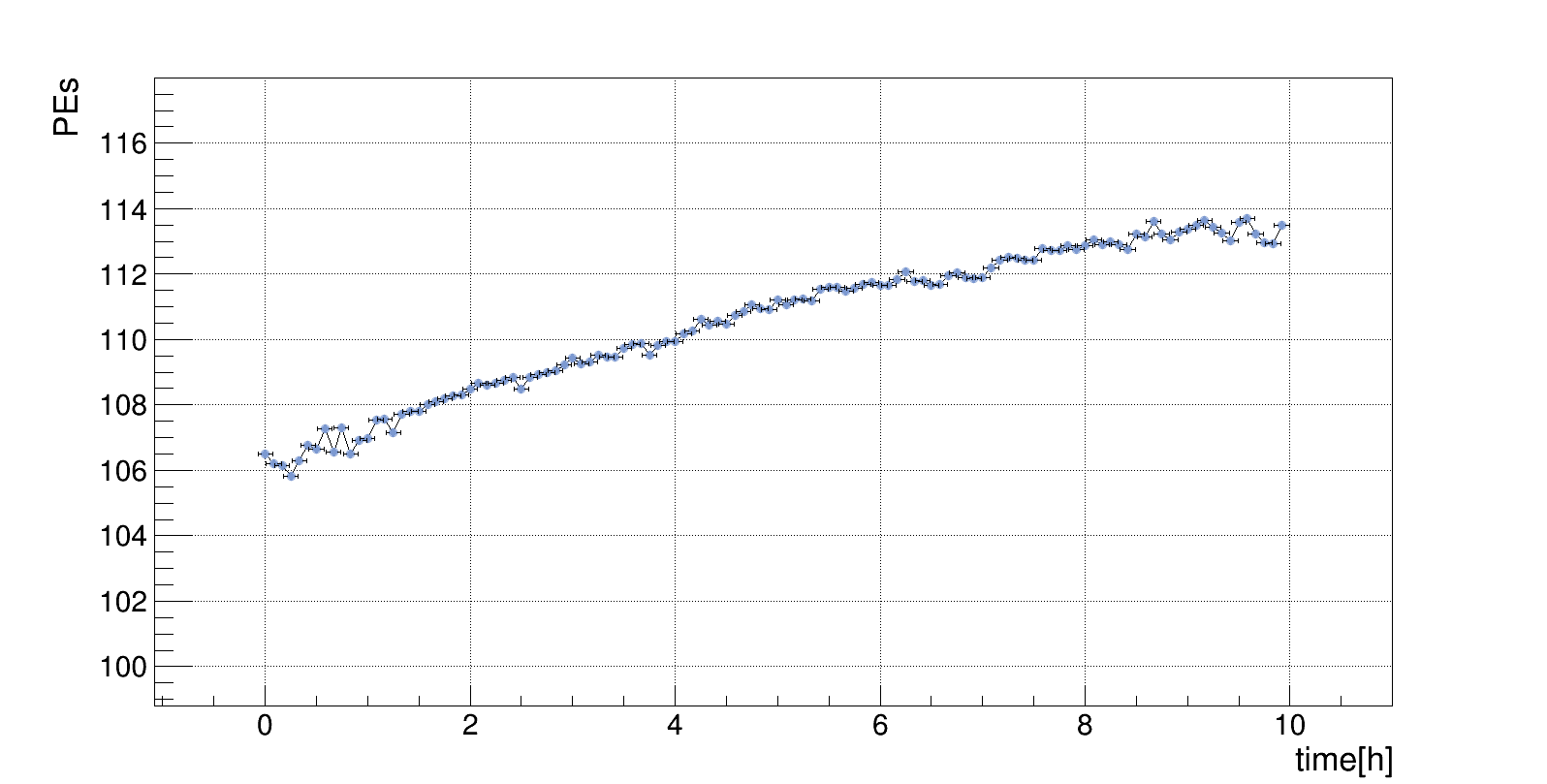}
\caption{Laser intensity monitoring in 10~hrs with a variation of 10$\%$}
\label{laserstability}
\end{figure*}

We have selected the laser, FQSS266-Q4-1k-STA~\cite{ref:GryLas}, which is a diode-pumped, passively Q-switched solid-state laser based on a
microchip laser with multiple frequency conversion stages. It is sealed in an aluminum box and the heat produced by the laser head is dissipated via the thermal conduction through a base plate. The maximum output power from the laser is 15~$\mu$J per pulse. It was found that the stability of the laser was correlated with ambient temperature. We use the air conditioner to control the ambient temperature to be 20$^{\circ}$C $\pm$ 1$^{\circ}$C. The stability of the laser measured by the monitor chamber (discussed later) in a typical 10~hrs is shown in Fig.~\ref{laserstability}.

\begin{figure*}
\centering
\includegraphics[width=1.0\textwidth]{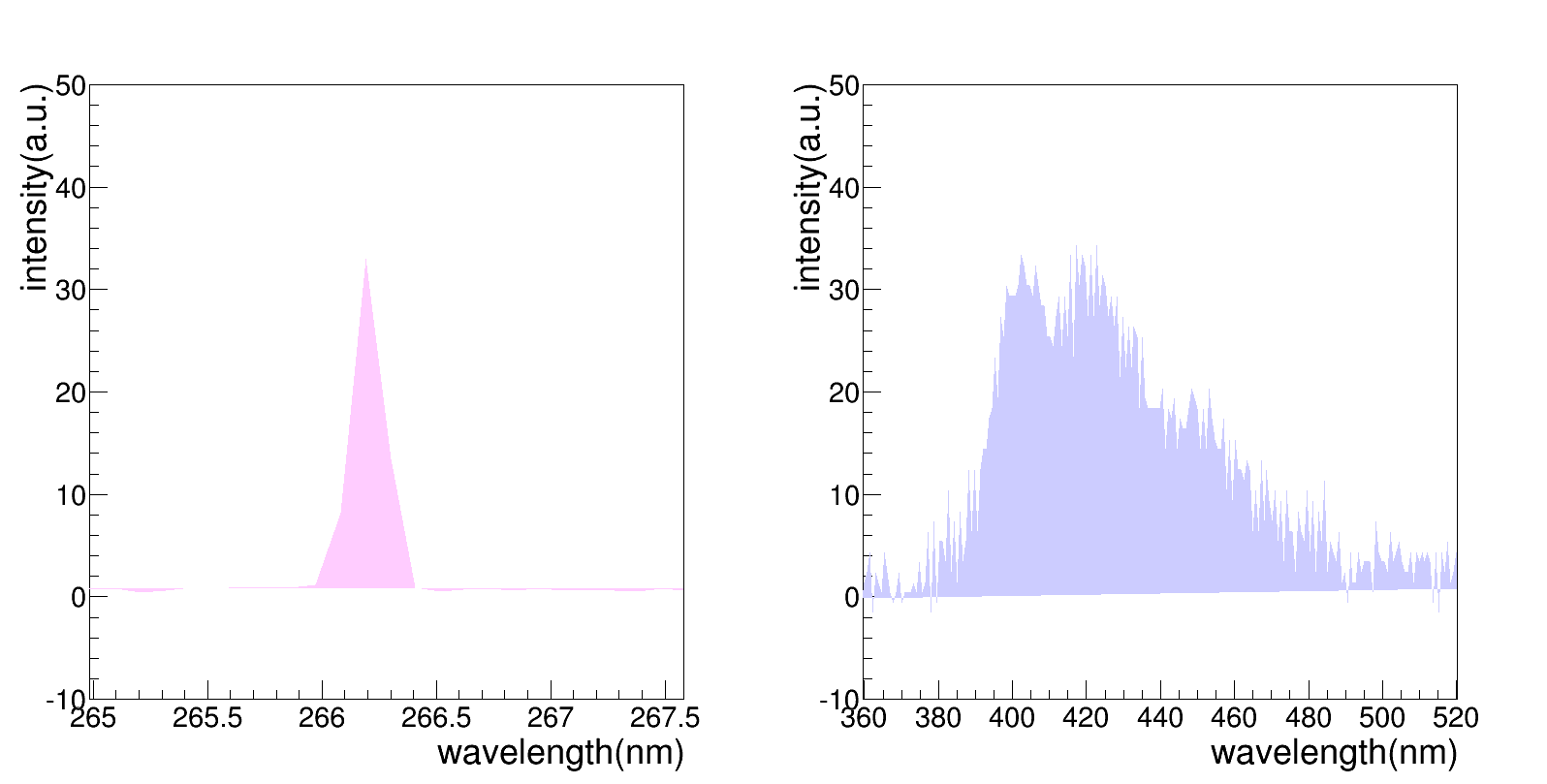}
\caption{Left figure: UV laser wavelength spectrum. Right figure: spectrum
  after the laser goes through the LS.}
\label{laserexciteLS}
\end{figure*}

The wavelength of the laser measured by a spectrometer (PG2000-Pro-EX, from Ideaoptics) is around 266~nm. After illuminating a sample of the JUNO LS, the resulting photon spectrum becomes the expected blue emission spectrum, as shown in Fig.~\ref{laserexciteLS}.

\subsubsection{Laser intensity adjustment}
The laser intensity can be adjusted either by internal control of the laser or by external optical filters in the beam. An electrical attenuator (a waveplate driven by a servo motor) designed for 266~nm inside the laser is capable of varying the intensity from 0$\%$ to 100$\%$. For the external control, motorized UV filter wheel FW102C~\cite{ref:thorlabs} is used, which allows six selectable optical attenuations of $10^{-0.1}(0.79)$, $10^{-0.2}(0.63)$, $10^{-0.3}(0.50)$, $10^{-0.4}(0.40)$, $10^{-0.5}(0.32)$, and $10^{-0.6}(0.25)$.

\subsubsection{Optical path}
To reduce the influence of the laser beam spot variation, a pinhole (P75H, from Thorlabs) is implemented on the beam path. A UV fused silica plate beam-splitter (BSW20, from Throlabs) is used to separates the laser beam into two (ratio: 50/50).  On each path, by a UV fused silica plano-convex lens, the laser is focused into an optical fiber held by a 3-D optical platform, which allows a range of 2~cm position adjustment with an accuracy of 0.01~mm.

\begin{figure*}[ht]
  \centering
  \subfigure[Fiber section(unit: mm)]{
  \includegraphics[width=2.6in]{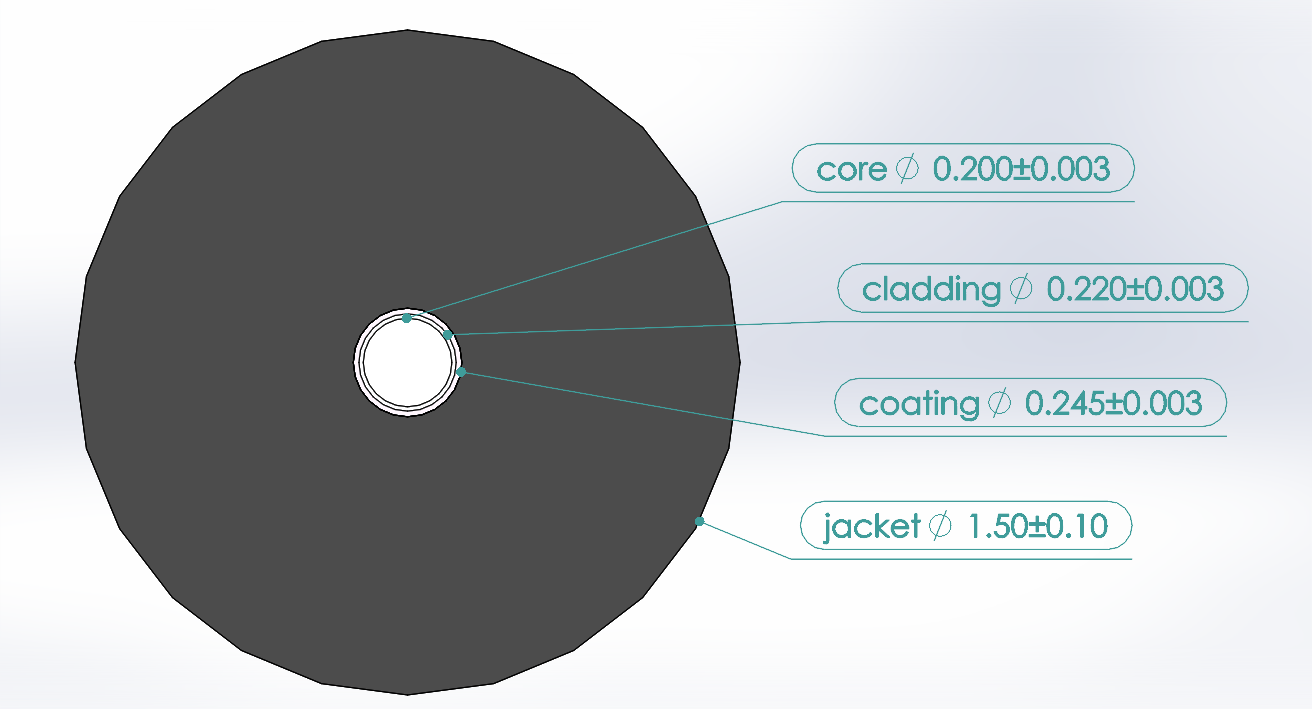}
  }
  \hspace{0.5in}
  \subfigure[Custom-made fiber]{
  \includegraphics[width=2.2in]{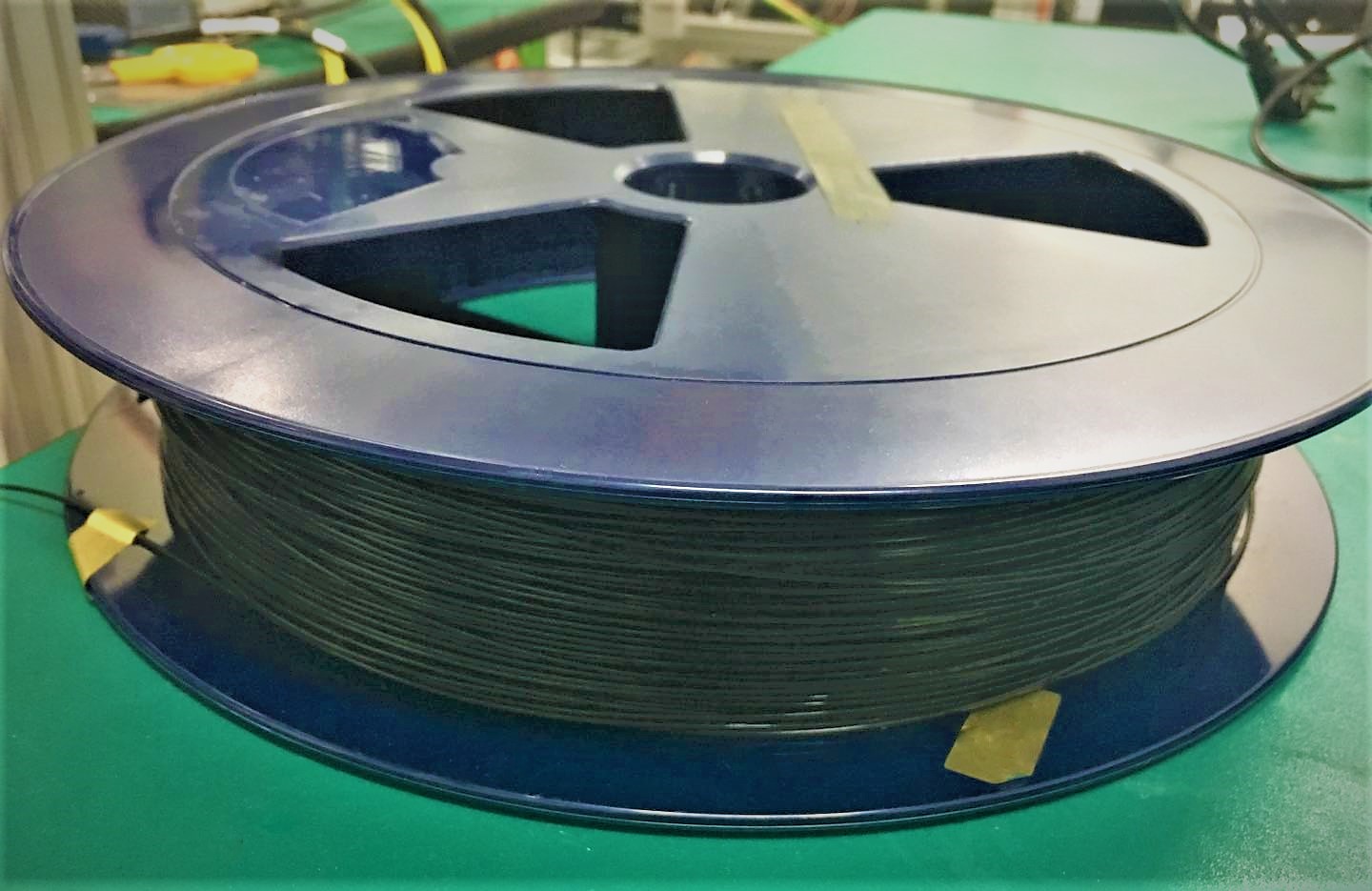}
  }
  \caption{Left figure: cross section of fiber. The core material
    is high hydroxyl quartz glass, the cladding material is F-doped
    silica glass, and coating material is bilayer UV-cured acrylic
    resin. The jacket of the fiber is ETFE. Right figure: the final fiber
    product.}
  \label{fiber}
\end{figure*}

The laser beam will be delivered into the central detector via a long optical fiber. This fiber has to be UV-transmitting, sufficiently thin and soft to be precisely spooled, shielded (photons emitted only through the open end), and compatible with the LS. We settled with multi-mode fibers with a core diameter of 200 $\mu$m and an external diameter of 1.5~mm (Fig.~\ref{fiber}), custom made by Beijing Scitlion Technology Corp., LTD (bare fiber and cladding) and Yangtze Optical Fibre and Cable Co., Ltd (jacket). The jacket is made with black ethylene-tetra-fluoro-ethylene (ETFE), and was tested to be compatible with the LS. The total light leakage through a 50~m fiber was measured to be less than 0.1\%.

\begin{figure*}[ht]
  \centering
  \includegraphics[width=5in]{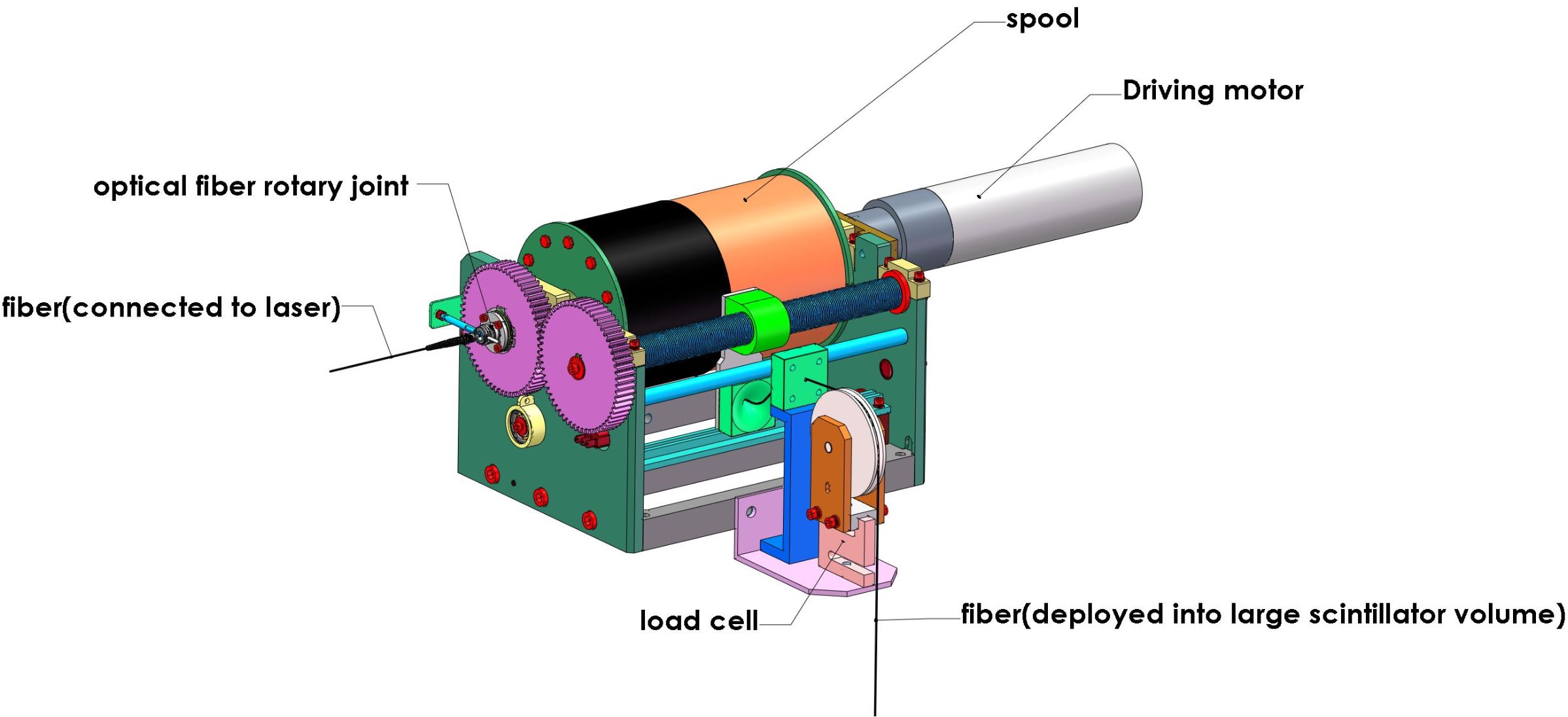}
  \caption{The deployment spool for the UV fiber}
  \label{fiberrotaryjoint}
\end{figure*}

To precisely deploy the fiber into and out of the central detector, a spool with a helical groove is developed. An optical fiber rotary joint (FO228, from MOOG) with a 200 $\mu$m bare UV fiber in the center~\cite{ref:fiberrotaryjoint} is implemented on the shaft of the spool to connect the 5~m fiber from the laser to the 50~m fiber on the spool. The UV photon reduction through this joint was measured to be 64$\%$.  A picture of the entire fiber motion assembly is shown in Fig.~\ref{fiberrotaryjoint}.

\subsubsection{Monitor and detector chambers}
The fibers described above enter into the detector and monitor chambers, then with their open end inserted into the scintillators. The wall of both chambers are covered by highly reflective PTFE reflectors to enhance the photon collection efficiency. Photons in each chamber are viewed by a 1-inch Hamamatsu R8520-406 PMT with excellent transit time jitter ($<$0.8~ns) and a 2-inch Hamamatsu R7725 PMT. The PMT signals are read out by a CAEN DT5720 digitizer (250~MS/s, 12-bit) for further analysis.

\section{System performance}
\label{sec:meas}
The laser pulses can be driven internally with a repetition rate between 1 to 1000~Hz. We performed tests at 50~Hz, the nominal rate for the JUNO laser calibration events. The digitizers are triggered by the synchronous laser driving pulses, and the waveform samples from the digitizers are integrated in a 40~ns window to obtain the total charge from each PMT.

\subsection{Photon intensity range in the detector chamber}
The PMT gains were monitored using the single photoelectrons (SPE) by low intensity laser runs. The typical gain was $2\times10^6$, and the stability was demonstrated to be stable in 0.15\% within 10~hrs, the duration of the intensity scan cycle (described later). The maximum number of scintillation photons from the liquid scintillator in the detector chamber is estimated to be $\sim$ $7 \times 10^{10}$ based on the PMT geometrical acceptance and quantum efficiency. Since the JUNO LS is expected to generate $\sim$10000 optical photons per MeV, the laser system can simulate up to a 7 TeV visible energy in the JUNO detector, sufficient even for showering muons.

\subsection{Light uniformity}
\begin{figure*}[htbp]
\centering
  \includegraphics[width=5.5in]{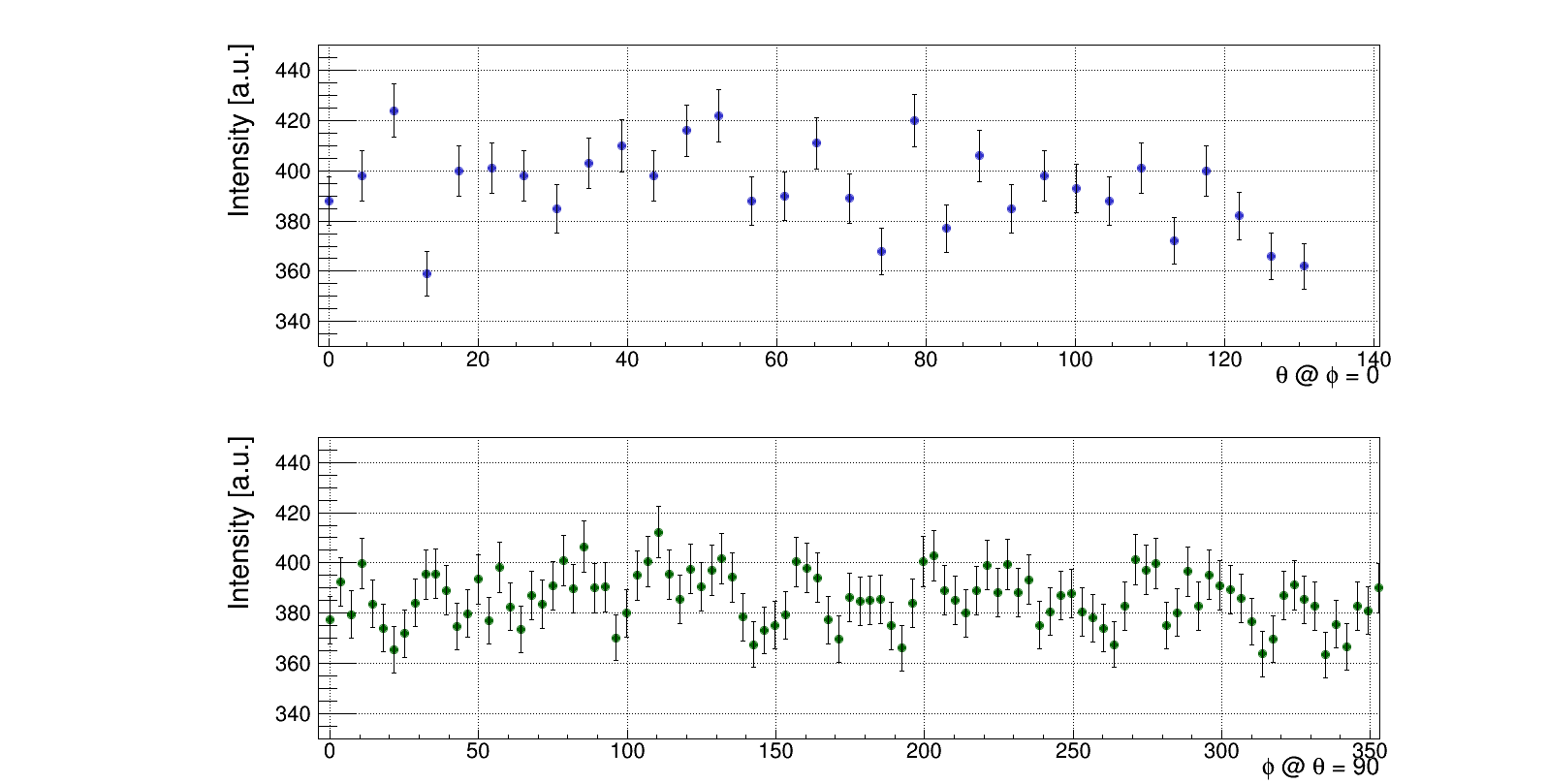}
  \caption{Photon intensity distribution in $\theta$ direction and $\phi$ direction.}
  \label{light_uniformity}
\end{figure*}

The isotropy of the laser light pulse injected into the large scintillator volume can be achieved by adding a diffuser ball to the open end of the 50~m fiber~\cite{ref:SNO_light_uniformity,ref:calib_SuperK}. We selected a solid PTFE sphere with a diameter of 15~mm and measured that the laser wavelength was unchanged after the diffusion. The best uniformity was achieved when the insertion depth of the fiber was 6.5~mm. As shown in Fig.~\ref{light_uniformity}, the variation of the angular distribution of the light intensity in the detection area is around 15$\%$ measured by the spectrometer. In reality, when the photons enter into the LS, the process of absorption and re-emission is expected to further smooth out the angular variation.

\begin{figure*}[htbp]
  \centering
  \includegraphics[width=4.5in]{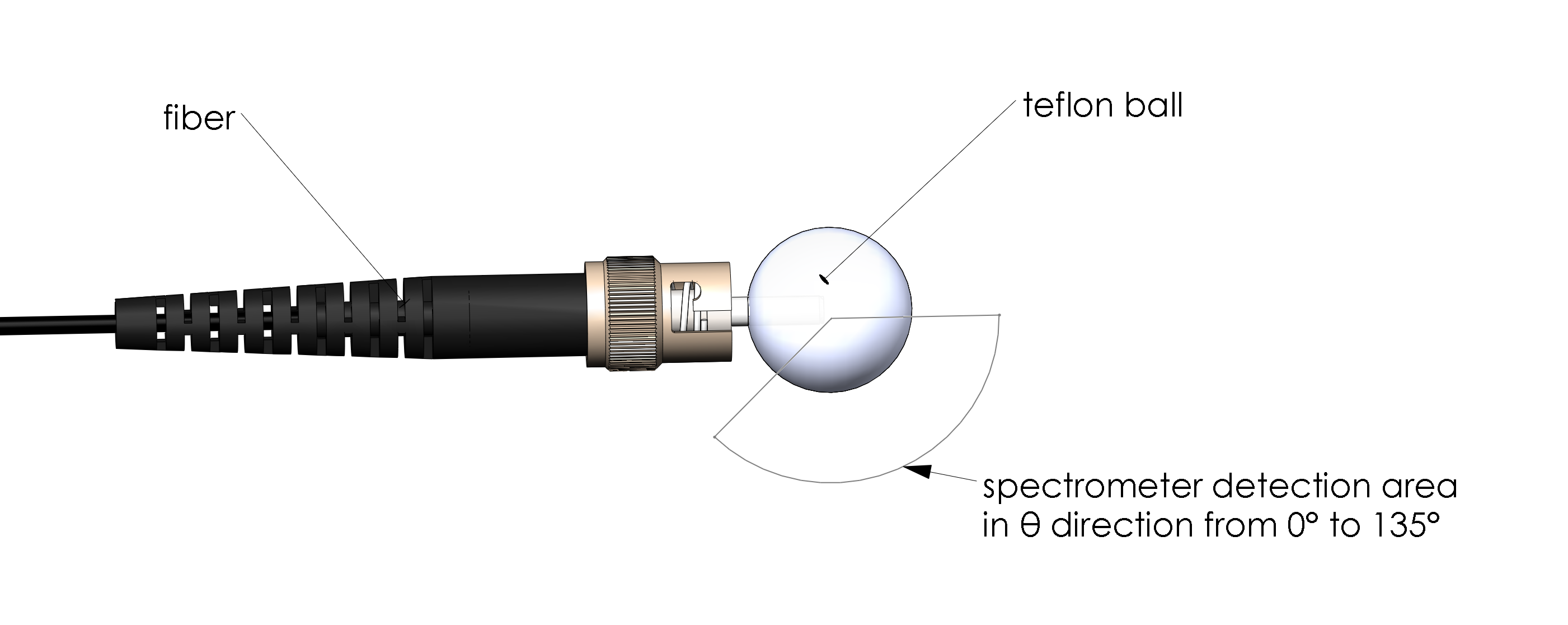}
  \caption{An illustration of the photon uniformity measurement}
\end{figure*}

\subsection{Precision of laser intensity monitoring}
Both the monitor and detector chambers contain two PMTs (different models and different linearities). As the laser intensity is varying up and down, in the perfect case, the ratio between the same-chamber PMTs would be a constant. Given that the two PMTs in the same chamber have different intrinsic nonlinearities, we use the standard deviation of this ratio over the range of beam intensity, $\sigma_{\rm same-chamber}$, to evaluate whether the PMTs are operated within their linear range. To monitor the laser intensity using the beam splitter, on the other hand, the stability of the splitting ratio may introduce an additional source of variation. Therefore, we used the standard deviation of the different-chamber ratio over the intensity range, $\sigma_{\rm diff-chamber}$, to estimate the overall nonlinearity of the monitoring system.

\begin{figure*}[htbp]
\centering
\includegraphics[width=5.5in]{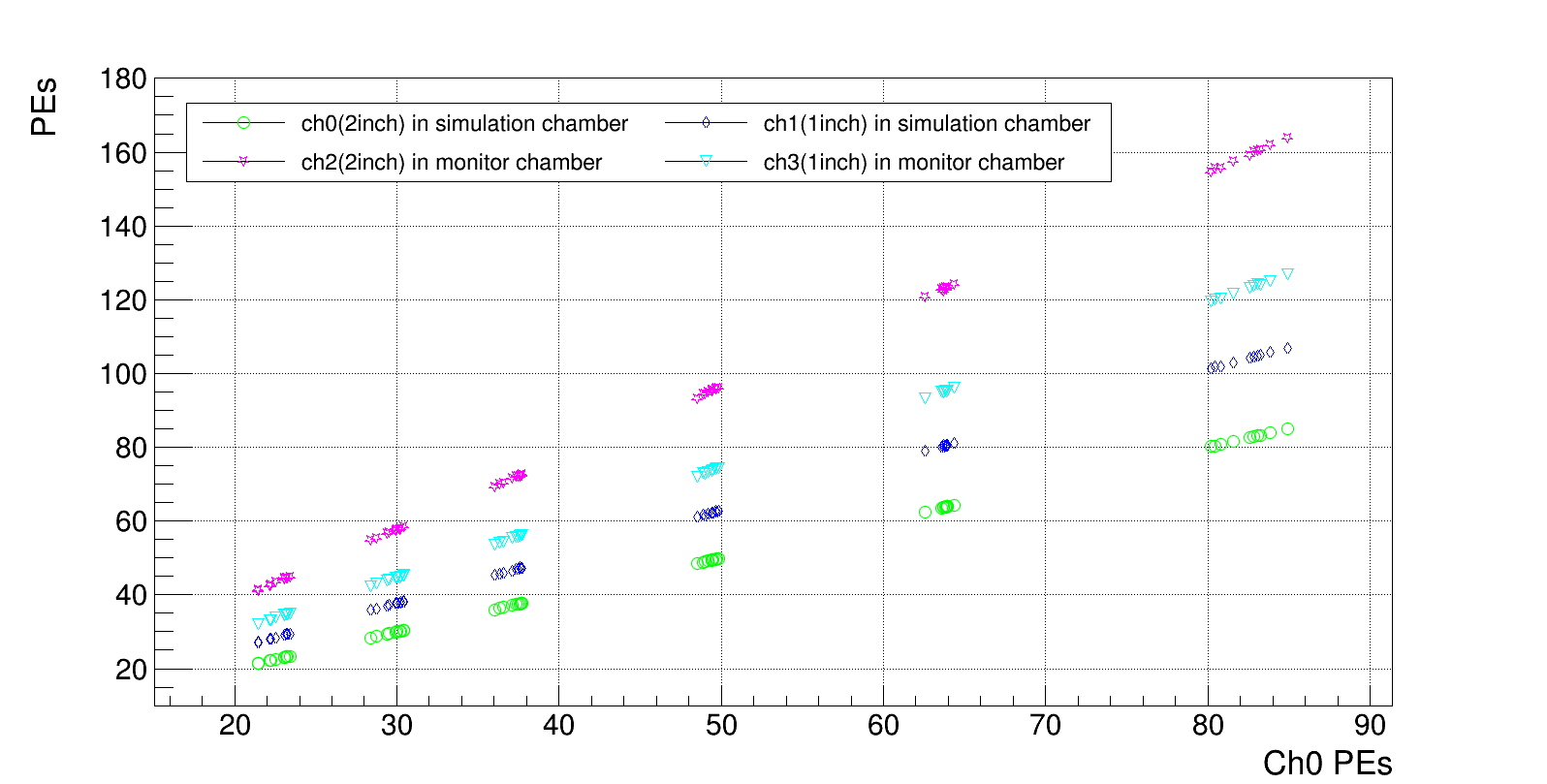}
\caption{Laser intensity test with PMTs: the slope for every channel should be constant along time.}
\label{sig0_lin}
\end{figure*}

In the JUNO detector, the primary energy of interest is between 1 to 8 MeV for the reactor neutrinos, where high precision measurement of the instrumental linearity is required. The total number of photons in this range is between 10000 to 80000, approximately. In our setup, to simulate this range of photons, the PMTs in both chambers should receive tens to a hundred PEs, assuming a geometrical acceptance of $1\%$ and $25\%$ quantum efficiency. After properly setting the attenuations, the laser intensity were scanned in 6 steps and repeatedly with a duration of about 10~hrs, as shown in Fig.~\ref{sig0_lin}.

\begin{figure*}[htbp]
\centering
\includegraphics[width=0.9\textwidth]{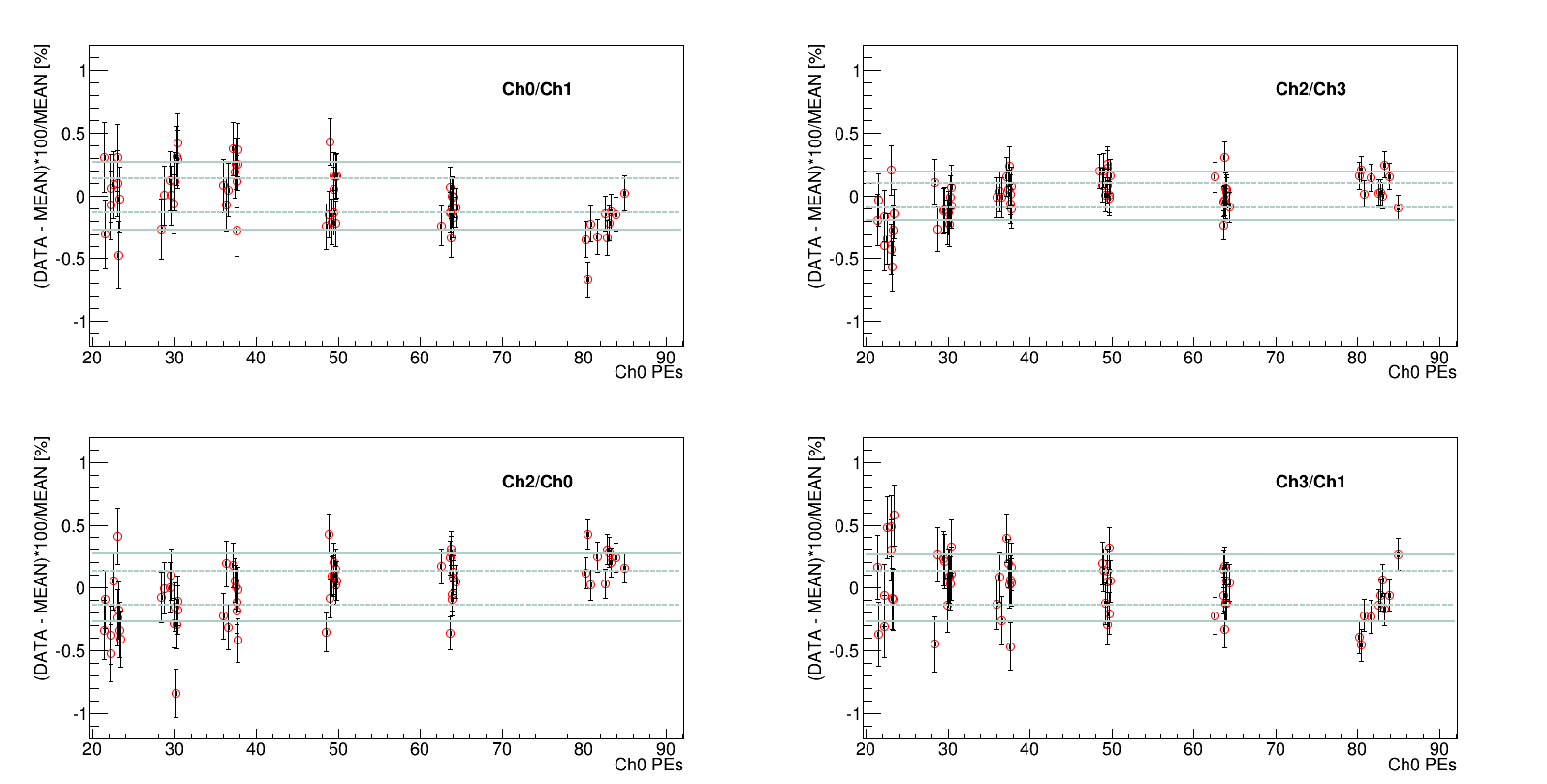}
\caption{Ratios between PMTs vs signal intensity of a given PMT (ch0). $\frac{ch0}{ch1}$ and $\frac{ch2}{ch3}$ are the same-chamber ratio
  for the monitor and detector chambers, respectively; $\frac{ch2}{ch0}$ and $\frac{ch3}{ch1}$ are the different-chamber
  ratios. The dashed line and the solid line give the 1-$\sigma_{\rm diff-chamber}$ and 2-$\sigma_{\rm diff-chamber}$ bands, respectively.}
\label{sig0_lin2}
\end{figure*}

The ratios of signals from the same-chamber PMTs and the different-chamber PMTs during the measurement period are shown in Fig.~\ref{sig0_lin2}. The same-chamber ratios appear to be quite stable, with $\sigma_{\rm same-chamber}$ $\sim$ 0.2\%. The variation of different-chamber's ratio appear to be slightly larger, with $\sigma_{\rm diff-chamber}$ in 0.3\%. Note that $\sigma_{\rm diff-chamber}$ combines both the variation in time as well as the nonlinearity of the intensity monitor. Conservatively, it can be used to quantify the accuracy of the laser intensity monitoring for this system.

\subsection{Laser induced photon timing}
\begin{figure*}[htbp]
\centering
\includegraphics[width=0.9\textwidth]{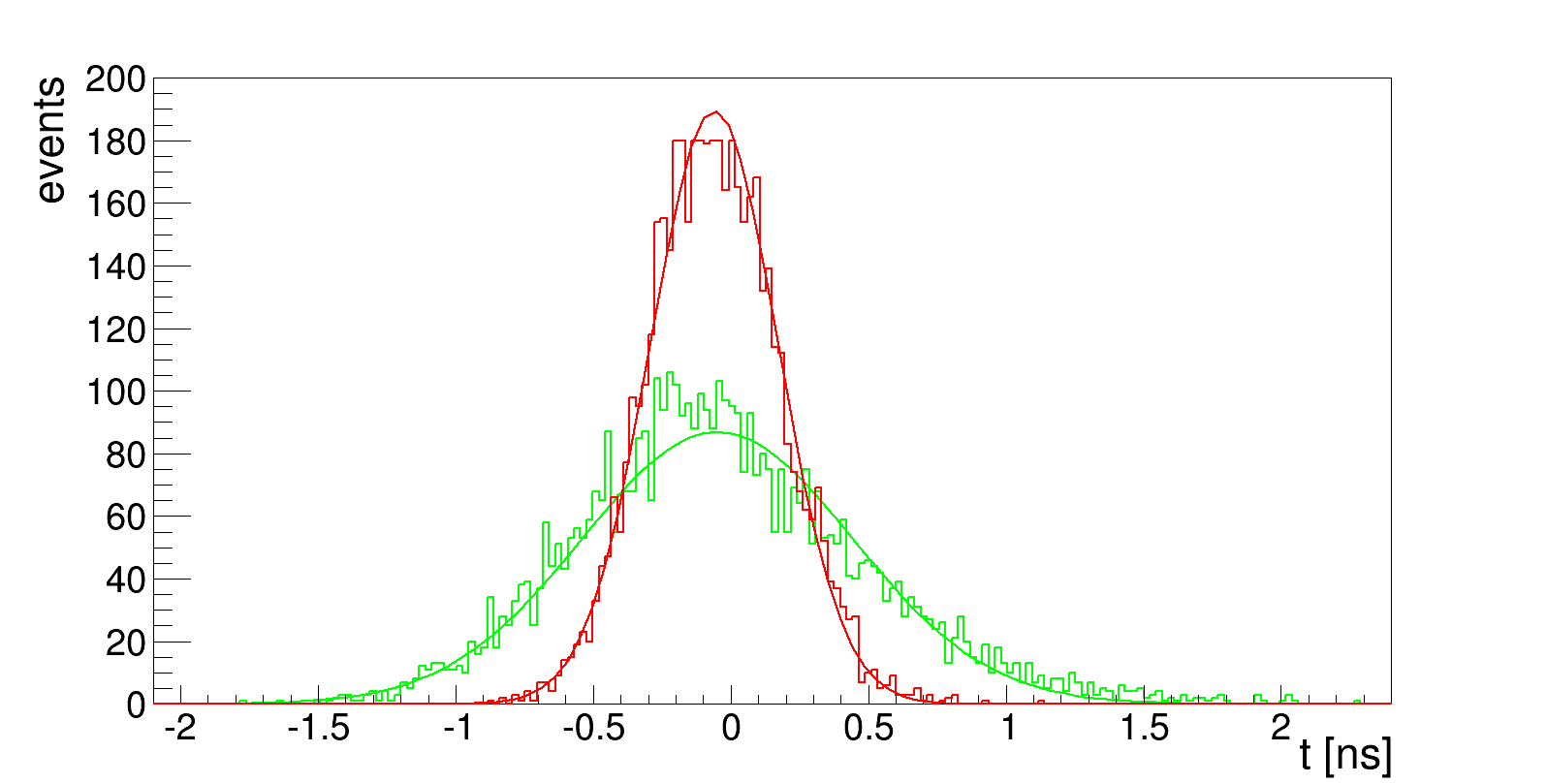}
\caption{Two histograms (centered at zero) showing the distribution of the rising edge in time when laser photons go directly into a PMT (red curve) or inject into the LS, then observed by a PMT (green curve).}
\label{LSexciting}
\end{figure*}

The timing of the PMTs and electronics can be calibrated by injecting the laser and aligning the rising edges of the detected pulses~\cite{ref:timing_calibration,ref:ALLAS_TCal2}. The bare laser timing has been checked by illuminating the UV photons directly onto the PMTs with an amplitude of $\sim$ 100~PEs. The distribution of the pulse rising edge is shown in Fig.~\ref{LSexciting}, where the standard deviation is 0.23~ns. The same study was performed by letting laser inject into a JUNO LS contained in an acrylic bottle. The PMT hit time was spread out, but the standard deviation is still within 0.5~ns. Therefore, the timing of the excited photons in the LS are still sufficiently fast to ensure a good timing calibration for the PMTs and electronics.

\section{Summary and conclusion}
\label{sec:con}
To summarize, a new UV laser calibration system has been developed for JUNO.  We discuss the performance of this system in this paper. The laser is capable of delivering photons up to an intensity of few TeV of equivalent deposited energy. The linearity of the intensity monitoring has been demonstrated to a 0.3\% level in a range suitable to calibrate the MeV-scale neutrino detector. The timing precision is significantly better than 1~ns. All these specifications satisfy the demanding needs for the JUNO experiment.

\section{Acknowledgement}
This work is supported by the Strategic Priority Research Program of the Chinese Academy of Sciences, Grant No. XDA10010800, and the CAS Center for Excellence in Particle Physics (CCEPP). We are thankful for the support from the Office of Science and Technology, Shanghai Municipal Government (Grant No. 16DZ2260200), and the support from the Key Laboratory for Particle Physics, Astrophysics and Cosmology, Ministry of Education.

\end{document}